\documentclass[%
reprint,
superscriptaddress,
amsmath,amssymb,
aps,
pre,longbibliography
]{revtex4-2}

\usepackage{graphicx}
\usepackage{dcolumn}
\usepackage{bm}

\usepackage{mathtools}

\usepackage{color}

\usepackage[hyperfootnotes=false]{hyperref}

\usepackage{graphicx}
\usepackage{float}

\usepackage[caption=false]{subfig}

\newtheorem{theorem}{Theorem}[section]
\newtheorem{remark}{Remark}[section]

\begin{document}
	
	\title{An Exactly Solvable Dual Bus Route Model}
	
	\author{Ngo Phuoc Nguyen Ngoc}
	\email[Corresponding author : ]{ngopnguyenngoc@duytan.edu.vn}
	\affiliation{Institute of Research and Development, Duy Tan University, Da Nang, 550000, Vietnam.}
	
	\affiliation{Faculty of Natural Sciences, Duy Tan University, Da Nang 550000, Vietnam.}

	\author{Huynh Anh Thi}
	\affiliation{Institute of Research and Development, Duy Tan University, Da Nang, 550000, Vietnam.}
	
	\affiliation{Faculty of Natural Sciences, Duy Tan University, Da Nang 550000, Vietnam.}
	
	\author{Nguyen Van Vinh}
	\affiliation{Faculty of Information Technology, Ho Chi Minh City University of Economics and Finance, Ho Chi Minh, 700000, Vietnam.}

	\begin{abstract}
		In 1998, O'Loan et al. introduced a simplified bus route model to illustrate bus dynamics. However, fluctuations in passenger numbers make it challenging to achieve an exact solution for the model's stationary state, as these fluctuations can impact bus behavior. In this study, we present an exactly solvable model for a dual bus route system that builds upon O'Loan et al.'s model. This dual model allows us to comprehensively analyze bus route dynamics. Our model introduces additional parameters not previously considered by O'Loan et al. to account for neighboring effects and align with an observation in the original model, which can influence the average stationary current and velocity of buses. When the neighboring effect is weak, our model behaves similarly to O'Loan et al.'s bus route model. However, with a strong neighboring effect, our model exhibits intriguing characteristics. Additionally, in some limiting cases, our model recovers well-known models such as the simple exclusion process, its generalization, the cooperative exclusion process, and the RNA polymerase model during the elongation stage.
	\end{abstract}
	\maketitle
	
	
	\section{Introduction}
	In a paper published in 1998, O’Loan et al. \cite{OLoan1998-1, OLoan1998-2} introduced a minimal bus route model (BRM) and its dual (DBRM). In this model, buses are treated as particles moving on a lattice, where the sites represent bus stops. The movement of the buses is heavily influenced by the presence of passengers at these stops. This model extends the Totally Asymmetric Simple Exclusion Process (TASEP) \cite{Liggett2005, Schutz2017, Spitzer1970}, originally proposed by Spitzer in 1970 to model exclusion processes inspired by biological traffic concerns \cite{MacDonald1968, MacDonald1969}.
	
	The bus route problem has also been explored using various other models, such as car-following models \cite{Hill2003, Huijberts2002, Nagatani2000, Nagatani2001, Nagatani2002-1} and nonlinear maps \cite{Nagatani2001-2}. Empirical studies on headway statistics in public transport systems, as analyzed in \cite{Krbalek2000}, align well with the Gaussian Unitary Ensemble of random matrix theory. Further models have been suggested in \cite{Baik2006, Krbalek2003}.
	
	A similar problem to the BRM is elevator traffic \cite{Nagatani2002, Nagatani2003, Nagatani2004}, where floors act as bus stops, often with varying inflow rates, particularly at the ground floor. BRM also resembles the ant-trail model proposed by Chowdhury et al. in 2002 \cite{Chowdhury2002, John2008, Schadschneider2005, Schadschneider2011}.
	In \cite{OLoan1998-2}, the authors delve into additional properties of the model and its dual. In the dual model, sites occupied by buses are considered empty. The model dynamics mimic the movements of both genuinely empty sites (neither a bus nor a passenger present) and sites with passengers, which can be seen as particles. The dual model is intriguing because the number of each type of particle can fluctuate, giving them degrees of freedom, which complicates finding an exact solution. Belitsky and Schütz \cite{Belitsky2019-1, Belitsky2019-2,Ngoc2024_2} addressed this challenge with a model for the pushing phenomenon in RNA polymerase, introducing a steady state accounting for fluctuating particle numbers. Following a similar approach, Ngoc et al. \cite{Ngoc2024} proposed an exactly solvable dual model of the ant-trail model \cite{Chowdhury2002}, facilitating its investigation. In this paper, we propose an exactly solvable model of the dual bus route model, akin to that in \cite{Ngoc2024}, but specifically tailored to the bus route dynamics. Let us first revisit the original bus route model and its dual.\\

\begin{figure}[h]
		\centering
		\includegraphics[width=0.41\textwidth]{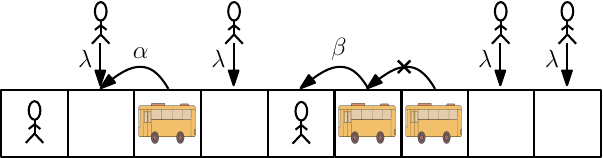}
		\caption{ Schematic illustration of dynamics of the BRM with all possible translations.}
		\label{brm_dynamics}
	\end{figure}

	\noindent \textbf{Bus route model (BRM):} The bus route is depicted as a lattice, with sites indexed clockwise from $1$ to $L$. At each site, there can be either a bus or a passenger. The dynamics of the BRM, illustrated in Fig. \ref{brm_dynamics}, follows a sequential random update rules:
	\begin{itemize}
		\item First, pick a site $i$ at random.
		\item If site $i$ is occupied by a passenger, nothing happens.
		\item If site $i$ is empty, a passenger arrives to occupy it with a rate of $\lambda$.
		\item If site $i$ is occupied by a bus, the bus jumps to the leftmost neighboring site (site $i-1$), provided that no other bus is present at the destination site. In this scenario, the jump rate depends on the presence of a passenger at site $i-1$. If a passenger is present, the rate is $\beta$; otherwise, it is $\alpha$.
	\end{itemize}
	
	In order to adapt to reality, a bus does not need to pick up passengers at a bus stop, it can travel faster. Therefore, we always assume that $\beta < \alpha$.\\
	
\begin{figure}[h]
		\centering
		\includegraphics[width=0.41\textwidth]{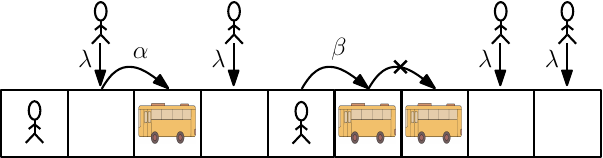}
		\caption{Schematic illustration of the dynamics of the DBRM with all possible translations.}
		\label{dual_brm_dynamics}
	\end{figure}

	\noindent\textbf{Dual bus route model (DBRM):} In the dual model, sites occupied by a bus are considered as empty sites. Thus, the motions can be seen as the movement of the real empty sites and sites occupied by a passenger. Therefore, the dynamics of the dual model, see Fig. \ref{dual_brm_dynamics},  is as follows:
	\begin{itemize}
		\item First, pick a site $i$ at random.
		\item If the site $i$ is occupied by a bus, nothing happens.
		\item If the site $i$ is empty, there are two possibilities:
		\begin{itemize}
			\item A passenger arrives to occupy it with a rate of $\lambda$.
			\item If the rightmost neighboring site (site $i+1$) is occupied by a bus, then the empty site ``jumps'' to that site with a rate of $\alpha$, i.e.,  site $i$ is now occupied by a bus and site $i+1$ becomes empty.
		\end{itemize}
		\item If site $i$ is occupied by a passenger, there are two possibilities:
		\begin{itemize}
			\item If site $i+1$ is not occupied by a bus, nothing happens.
			\item If the rightmost neighboring site is occupied by a bus, the bus moves to site $i$ with a rate of $\beta$, picking up the passenger. Thus, site $i$ is now occupied by the bus, and site $i+1$ becomes empty.
		\end{itemize}
	\end{itemize}

	As mentioned earlier, the fluctuation in passenger numbers makes it challenging to find an exact solution for the steady state. Mean-field theory or mapping the processes to a zero-range process can be employed to analyze the models, as discussed in \cite{OLoan1998-1, OLoan1998-2}.\\
	
	\noindent\textbf{Organization:} The remainder of the paper is structured as follows: In Section \ref{model}, we introduce the model, including its dynamics and the stationary distribution in parameterized forms. We then outline the conditions on the parameters necessary to align the dynamics with the stationary measure. Section \ref{current_velocity} explores the average stationary current and velocity of particles within our dual bus route model and extends these findings to buses. We proceed to discuss the results in Section \ref{discussion}. Finally, in Section \ref{Summary}, we summarize the findings of this work.
	
	\section{Dual bus route model with short-range interaction}\label{model}
	\subsection{The model state space and dynamics}
	We consider the bus route as a lattice, where each site represents a bus stop. At each site, there can be either a bus or a passenger. In our model, sites on the bus route that are occupied by a bus are treated as empty sites. Thus, a site with a bus is designated as state 0, representing an empty location. A site without a bus but with a passenger is denoted as state 1, while a genuinely unoccupied site is labeled as state 2. This setup bears resemblance to the one described in \cite{Ngoc2024}, where ants and pheromones play roles analogous to buses and passengers in our model. In that study, a site occupied by an ant is denoted by 0, mirroring our model. However, unlike our approach, in that work, an empty site is denoted by 1 rather than 2. With this setup, the dynamics of the original model, i.e., O’Loan et al.'s model, are simply expressed as
	\begin{equation}\label{simple_dynamics}
		\begin{cases}
			20 \to 02,\ \ \text{with rate } \alpha,\\
			10 \to 02,\ \ \text{with rate } \beta,\\
			2 \to 1,\ \ \text{with rate } \lambda.
		\end{cases}
	\end{equation}
	It is worth noting that the dynamics described above are identical to those presented in \cite{Ngoc2024} in terms of symbolism.

	To elucidate the dynamics of our model, we first define its state space. Given our focus on the collective movement of particles in DBRM, including buses in BRM, we adopt periodic boundary conditions. This implies that the lattice, defined as $\mathbb{T}_L := \mathbb{Z}/L\mathbb{Z}$, consists of $L$ sites, where $i+L \equiv i \mod L$. Each configuration of particles within this lattice is represented by $\boldsymbol{\eta} = (\eta_1, \ldots, \eta_L)$, where local variables $\eta_j \in \{0, 1, 2\}$ for $j = 1, 2, \ldots, L$ representing the state of site $j$.
	
	\begin{figure}[h]
		\centering
		\includegraphics[width=0.25\textwidth]{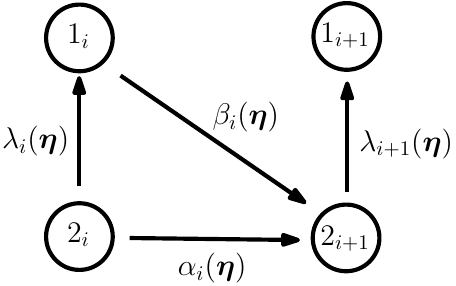}
		\caption{Minimal reaction scheme for particle translocation: A particle can move from site $i$ to site $i+1$ if it is in state 1 (indicating a site with a passenger) and changes its state to 2 (representing a "real" empty site) with an effective rate of $\beta_i(\boldsymbol{\eta})$. Similarly, if the particle is in state 2 at site $i$, it can move to site $i+1$ at a rate of $\alpha_i(\boldsymbol{\eta})$, maintaining its state. Additionally, at site $i$, a particle in state 2 can change to state 1 without moving, indicating a passenger arriving at the bus stop. This transition occurs at a rate of $\lambda_i(\boldsymbol{\eta})$.}
		\label{single_bus_dynamics}
	\end{figure}
	
	In this work, we assume that the transition rates depend on the surrounding environment, specifically the presence of buses and passengers. Therefore, in our framework, the rates of a particle are contingent upon the current positions of buses and passengers. Consequently, the rates $\alpha, \beta$, and $\lambda$ are now denoted as $\alpha_i(\boldsymbol{\eta})$, $\beta_i(\boldsymbol{\eta})$, and $\lambda_i(\boldsymbol{\eta})$ for a particle $i$ to elucidate the dependence of transition rates on the current configuration of the system. For an illustration of the dynamics involving a single particle for both types, refer to Fig. \ref{single_bus_dynamics}.

	It is evident that an allowed configuration, denoted by $\boldsymbol{\eta} = \{\eta_1, \ldots, \eta_L\}$, can be described by the positions and states of particles (1 and 2). Let us consider there are $N$ particles of both types 1 and 2. This implies that there are $L-N$ empty sites on the lattice, corresponding to $L-N$ buses. We denote the position and state vectors of particles as $\textbf{x} = (x_1, \ldots, x_N)$ and $\textbf{s} = (s_1, \ldots, s_N)$ with $s_i \in \{1,2\}$, respectively.
	
	Due to periodicity, the positions $x_i$ of the particles are counted modulo $L$, and the labels $i$ are counted modulo $N$. Utilizing the position vector offers the advantage of expressing the distance between two particles using the Kronecker-$\delta$ function, defined as follows:
	\begin{equation}
		\delta_{k,\ell} =
		\begin{cases}
			1, & \text{if } k = \ell,\\
			0, & \text{if } k \neq \ell,
		\end{cases}
	\end{equation}
	for any $k,\ell$ belonging to any set. For example, to indicate that the $i^{th}$ and $(i+1)^{th}$ particles are located at neighboring sites, one can use the notation $\delta_{x_i+1,x_{i+1}}$. This function evaluates to 1 if the two particles are indeed located at neighboring sites, and 0 otherwise.
	
	Now, we can introduce the rates influenced by the surrounding environment. Let us consider the $i^{th}$ particle, which can make a rightward jump with rates $\alpha_i(\boldsymbol{\eta})$ and $\beta_i(\boldsymbol{\eta})$, contingent on whether its state is 2 or 1, respectively. Additionally, if its state is 2, there exists another possibility for transitioning to state 1, governed by the rate $\lambda_i(\boldsymbol{\eta})$. The rates of the models are given by:
	
	\begin{align}
		\alpha_{i}(\boldsymbol{\eta}) =\ & \delta_{s_{i},2}\alpha^{\star}(1+\delta_{x_{i-1},x_{i}-1}\alpha^{1\star})(1-\delta_{x_{i}+1,x_{i+1}}),\label{rate_1}\\
		\beta_{i}(\boldsymbol{\eta})  =\  & \delta_{s_{i},1}\beta^{\star}(1+\delta_{x_{i-1},x_{i}-1}\beta^{1\star})(1-\delta_{x_{i}+1,x_{i+1}}),\label{rate_2}\\
		\lambda_{i}(\boldsymbol{\eta}) 	=\  &
		\delta_{s_{i},2}\lambda^{\star}(1+\delta_{x_{i-1},x_{i}-1}\lambda^{1\star}
		+ \delta_{x_{i}+1,x_{i+1}}\lambda^{\star1}  \nonumber\\
		& + \lambda^{1\star1}\delta_{x_{i-1},x_{i}-1}\delta_{x_{i}+1,x_{i+1}}).\label{rate_3}
	\end{align}
	
\begin{figure}
		\subfloat[$\alpha_i(\boldsymbol{\eta}) = \alpha^\star$\label{jump_rate_alpha_a}]{\includegraphics[width=3.2cm]{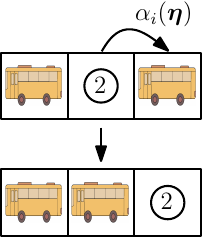}}
		\hspace{0.5cm}
		\subfloat[$\alpha_i(\boldsymbol{\eta}) = \alpha^\star(1 + \alpha^{1\star})$\label{jump_rate_alpha_b}]{\includegraphics[width=3.2cm]{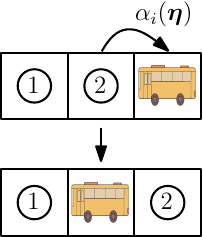}}\\
		\caption{Here are two illustrations depicting the translocation rate $\alpha_i(\boldsymbol{\eta})$ for a particle in state 2. In the left panel, the rate is $\alpha^\star$ because there is no contribution from $\alpha^{1\star}$ since there is no particle present on the left. In contrast, in the right panel, there is a particle on the left of the considered particle, resulting in a rate of $\alpha^\star(1 + \alpha^{1\star})$.}
		\label{jump_rate_alpha}
	\end{figure}

\begin{figure}
		\centering
		\subfloat[$\beta_i(\boldsymbol{\eta}) = \beta^\star$\label{jump_rate_beta_a}]{\includegraphics[width=3.2cm]{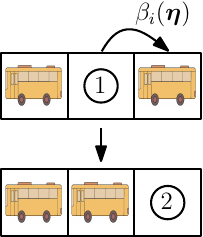}}
		\hspace{0.5cm}
		\subfloat[$\beta_i(\boldsymbol{\eta}) = \beta^\star(1 + \beta^{1\star})$\label{jump_rate_beta_b}]{\includegraphics[width=3.2cm]{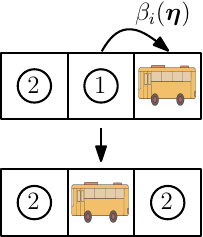}}\\
		\caption{Here are two illustrations depicting the translocation rate $\beta_i(\boldsymbol{\eta})$ for a particle in state 1. In the left panel, the rate is $\beta^\star$ because there is no contribution from $\beta^{1\star}$ since there is no particle present on the left. In contrast, in the right panel, there is a particle on the left of the considered particle, resulting in a rate of $\beta^\star(1 + \beta^{1\star})$. }
		\label{jump_rate_beta}
	\end{figure}

\begin{figure}
		\subfloat[$\lambda_i(\boldsymbol{\eta}) = \lambda^\star$ \label{transition_rate_lambda_a}]{\includegraphics[width=3.2cm]{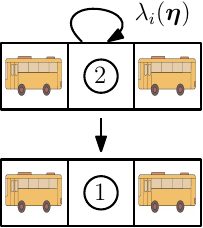}}
		\hspace{0.5cm}
		\subfloat[$\lambda_i(\boldsymbol{\eta}) = \lambda^\star(1 + \lambda^{\star1})$\label{transition_rate_lambda_b}]{\includegraphics[width=3.2cm]{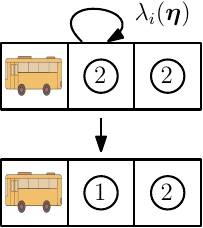}}\\
		\caption{Here are two illustrations depicting the translocation rate $\lambda_i(\boldsymbol{\eta})$ for a particle in state 2. In the left panel, the rate is $\lambda^\star$ since there are no particles present on both sides of the particle. In the right panel, while there is a particle on the right but not on the left, only $\lambda^{\star1}$ contributes to the rate, not $\lambda^{1\star}$. Thus, the rate in this case is $\lambda^\star(1 + \lambda^{\star1})$.}
		\label{transition_rate_lambda}
	\end{figure}

	Before delving into the meaning of each term in the rates, it is important to clarify that the dynamics governed by these rates differ from those in Ngoc et al. \cite{Ngoc2024}. Further explanations are provided in Remark \ref{diff_dynamics}. Notably, the parameters $\alpha^\star$, $\beta^\star$, and $\lambda^\star$ play similar roles to $\alpha$, $\beta$, and $\lambda$ in O'Loan et al.'s model. If one sets the other parameters to 0, i.e., $\alpha^{1\star} = \beta^{1\star} = \lambda^{1\star} = \lambda^{\star1} = \lambda^{1\star1} = 0$, the rates $\alpha_{i}(\boldsymbol{\eta})$, $\beta_{i}(\boldsymbol{\eta})$, and $\lambda_{i}(\boldsymbol{\eta})$ become $\alpha^\star$, $\beta^\star$, and $\lambda^\star$ respectively, thus recovering the original model. For visual representations of rates \eqref{rate_1}, \eqref{rate_2}, and \eqref{rate_3}, see Figs. \ref{jump_rate_alpha}, \ref{jump_rate_beta}, and \ref{transition_rate_lambda}, respectively.
	
	In equations \eqref{rate_1} and \eqref{rate_2}, the term $(1-\delta_{x_{i}+1,x_{i+1}})$ indicates that the effectiveness of the jump rates $\alpha_{i}(\boldsymbol{\eta})$ and $\beta_{i}(\boldsymbol{\eta})$ depends on whether the next site to the right of particle $i$ is unoccupied. The superscript $\star$ designates the position of the considered particle $i$, with symbols 0 and 1 representing unoccupied and occupied sites, respectively. Consequently, the parameters $\alpha^{1\star}$ and $\beta^{1\star}$ represent the contribution to $\alpha_{i}(\boldsymbol{\eta})$ and $\beta_{i}(\boldsymbol{\eta})$, respectively, when the $i^{th}$ and $(i-1)^{th}$ particles are positioned at adjacent sites, indicating the absence of a bus at site $x_{i}-1$. The condition of zero distance between the two particles is indicated by $\delta_{x_{i-1}, x_{i}-1}$, which equals 1 only when the two particles are indeed located at neighboring sites.

	\textbf{An observation in O'Loan et al.'s BRM:} When the distance between buses increases due to fluctuations, it tends to widen further: since buses move faster in less populated areas, a bus following closely behind another is likely to accelerate more than one that is farther behind the lead bus. This occurs because the closer a bus is to the one directly ahead of it, passengers will have had less time to arrive.
	
	\textbf{Neighboring-effect parameters $\alpha^{1\star}$ and $\beta^{1\star}$:} Let us clarify the role of the parameter $\alpha^{1\star}$ using Fig. \ref{jump_rate_alpha} as illustration. In Fig. \ref{jump_rate_alpha_a}, we observe a scenario where the particle type 2 has two neighbors are empty sites (occupied by buses), resulting in a jump rate of the particle with state 2 being $\alpha(\boldsymbol{\eta}) = \alpha^\star$. However, when there is no bus at the leftmost neighboring site of the particle, i.e., there is the presence of a particle at this site, as shown in Fig. \ref{jump_rate_alpha_b}, the rate becomes $\alpha(\boldsymbol{\eta}) = \alpha^\star(1+ \alpha^{1\star})$. Therefore, selecting $\alpha^{1\star} < 0$ indicates that the latter rate is lower than the former. This choice aligns with the earlier observation that a bus moves slower when it is further from the leading bus. The role of $\beta^{1\star}$ mirrors that of $\alpha^{1\star}$, using Fig. \ref{jump_rate_beta} as illustration. Hence, we designate both parameters $\alpha^{1\star}$ and $\beta^{1\star}$ as neighboring-effect parameters, given their influence on the rates at which buses locate neighboring sites.

	\begin{remark}\label{diff_dynamics}
		Although the minimal reaction scheme of particle translocation in Fig. \ref{single_bus_dynamics} is the same as that in \cite{Ngoc2024}, the jump rates \eqref{rate_1}-\eqref{rate_2} in this model differ from those in \cite{Ngoc2024}. Specifically, the particle jump rates in this model are influenced by the presence of a particle directly behind it, whereas in the model presented in \cite{Ngoc2024}, the jump rates of a particle depend on the presence of a particle at the next-nearest site in front of it.
	\end{remark}

	\subsection{The model stationary distribution}
	Recall that a configuration $\boldsymbol{\eta}$ of the system can be specified by the position and state vectors $\textbf{x} = (x_1,x_2,...,x_N)$ and $\textbf{s} = (s_1,s_2,...,s_N)$. The probability of finding the system in configuration $\boldsymbol{\eta}$ in equilibrium is assumed to follow this form:
	
	\begin{equation}\label{inv_meas}
		\pi(\boldsymbol{\eta})=\dfrac{1}{Z}e^{-\frac{1}{k_{B}T}(U(\textbf{x})+\lambda B(\textbf{s}))}.
	\end{equation}
	Here, firstly, $T$ plays a role similar to temperature, and $k_B$ represents the Boltzmann constant. Secondly, $U$ represents the short-range interaction energy, defined as
	\begin{equation}
		U(\textbf{x})=J\sum_{i=1}^{N}\delta_{x_{i}+1,x_{i+1}}.
	\end{equation}
	Positive
	$J$ corresponds to repulsion. Thirdly, $B$ denotes the excess, calculated as $N^{1}-N^{2}$, where $N^{\alpha}$ represents the fluctuating number of sites in state $\alpha \in \{1,2\}$. One has $N^{1} + N^{2} = N$ and 
	\begin{equation}
		B(\textbf{s})=\sum_{i=1}^{N}(3-2s_{i}).
	\end{equation}
	The chemical potential $\lambda$ is a Lagrange multiplier that takes care of the fluctuations in the excess due to the interplay of transition state of the particles. Finally, $Z$ is the partition function.   
	
	The form of the invariant measure \eqref{inv_meas} mirrors those in \cite{Belitsky2019-1,Belitsky2019-2} and \cite{Ngoc2024}. However, it is important to note that the values of parameters in our measure will differ from those in those papers due to differences in our model's dynamics.
	
	For the convenience of computation, one introduces
	
	\begin{equation}
		x=e^{\frac{2\lambda}{k_{B}T}}\text{ and }y=e^{\frac{J}{k_{B}T}},
	\end{equation}
	so that $x>1$ corresponds to an excess of particles in state 1 and repulsive interaction corresponds to $y>1$.
	
	To ensure that the process, guided by the dynamics \eqref{rate_1}--\eqref{rate_3}, possesses a measure in the form of \eqref{inv_meas} as its invariant distribution, a prerequisite is that the model's parameters must adhere to the following constraints.
	
	\begin{align}
		x&=\dfrac{\beta^{\star}}{\lambda^\star}\label{cond1}\\
		y&=\dfrac{1+\beta^{1\star}+\dfrac{1}{\lambda^{\star}}\alpha^{\star}(1+\alpha^{1\star})}{1+\dfrac{1}{\lambda^{\star}}\alpha^{\star}}\\
		\lambda^{1\star}&=\dfrac{x}{1+x}(1+\beta^{1\star})-\dfrac{x}{1+x}\dfrac{\alpha^{\star}}{\beta^{\star}}(1+\alpha^{1\star})-1\\
		\lambda^{\star1}&=\dfrac{1}{1+x}(1+\beta^{1\star})+\dfrac{x}{1+x}\dfrac{\alpha^{\star}}{\beta^{\star}}(1+\alpha^{1\star})-1\\
		\lambda^{1\star1} & = -\beta^{1\star}\label{cond2}
	\end{align}
	
	In other words, one can summarize the result by Theorem \ref{thm} which is the main result of the present work.
	
	\begin{widetext}
		
		\begin{theorem}\label{thm}
			If the parameters appearing in the rates \eqref{rate_1}--\eqref{rate_3} satisfy constraints \eqref{cond1}--\eqref{cond2}, then the 	invariant measure of the process is the following
			\begin{equation}\label{main_theorem}
				\pi(\boldsymbol{\eta}) = \frac{1}{Z}\left(\dfrac{\beta^{\star}}{\lambda^{\text{\ensuremath{\star}}}}\right)^{\sum_{i=1}^{N}-3/2+ s_i}\left( \frac{1+\beta^{1\star}+\dfrac{1}{\lambda^{\star}}\alpha^{\star}(1+\alpha^{1\star})}{1+\dfrac{1}{\lambda^{\star}}\alpha^{\star}} \right)^{-\sum_{i=1}^{N}\delta_{x_{i+1}, x_i+1}}
			\end{equation}
			where $Z$ is the partition function.
		\end{theorem}
	\end{widetext}
	
	The proof of the theorem closely resembles that presented in \cite{Ngoc2024}. Therefore, we will only provide a brief outline of the proof. At equilibrium, the master equation governing the probability of the system being in configuration $\boldsymbol{\eta}$ is as follows: 
	\begin{equation}\label{master_eq}
		\begin{split}
			\sum_{i=1}^{N}&\bigg(\alpha_{i}(\boldsymbol{\eta}_{1}^{i})\pi(\boldsymbol{\eta}_{1}^{i})+\beta_{i}(\boldsymbol{\eta}_{2}^{i})\pi(\boldsymbol{\eta}_{2}^{i})+\lambda_{i}(\boldsymbol{\eta}_{3}^{i})\pi(\boldsymbol{\eta}_{3}^{i})\\
			&-(\alpha_{i}(\boldsymbol{\eta})+\beta_{i}(\boldsymbol{\eta})+\lambda_{i}(\boldsymbol{\eta}))\pi(\boldsymbol{\eta})\bigg) = 0.
		\end{split}
	\end{equation}
	Here, $\boldsymbol{\eta}_{1}^{i}$ represents the configuration before the forward translocation of particle $i$ in state 2, leading to $\boldsymbol{\eta}$ (with $x_{i}^{1}=x_{i}-1$ and $s_{i}^{1}=s_{i}$); $\boldsymbol{\eta}_{2}^{i}$ signifies the configuration before the forward translocation of particle $i$ in state 1, leading to $\boldsymbol{\eta}$ ($x_{i}^{2}=x_{i}-1$ and $s_{i}^{2}=3-s_{i}$);  $\boldsymbol{\eta}_{3}^{i}$ denotes the configuration before a passenger arrives at site $x_i$, leading to $\boldsymbol{\eta}$ ($x_{i}^{3}=x_{i}$ and $s_{i}^{3}=3-s_{i}$). 
	
	Dividing both sides of \eqref{master_eq}  by  $\pi(\boldsymbol{\eta})$, one obtains
	\begin{equation}
		\begin{split}
			\sum_{i=1}^{N}\bigg(\alpha_{i}(\boldsymbol{\eta}_{1}^{i})&\dfrac{\pi(\boldsymbol{\eta}_{1}^{i})}{\pi(\boldsymbol{\eta})}+\beta_{i}(\boldsymbol{\eta}_{2}^{i})\dfrac{\pi(\boldsymbol{\eta}_{2}^{i})}{\pi(\boldsymbol{\eta})}+\lambda_{i}(\boldsymbol{\eta}_{3}^{i})\dfrac{\pi(\boldsymbol{\eta}_{3}^{i})}{\pi(\boldsymbol{\eta})}\\
			&-(\alpha_{i}(\boldsymbol{\eta})+\beta_{i}(\boldsymbol{\eta})+\lambda_{i}(\boldsymbol{\eta}))\bigg)=0.\label{mas_eq2}
		\end{split}
	\end{equation}
	Let us denote the following quantities
	\begin{align}
		A_{i}(\boldsymbol{\eta)} & =\alpha_{i}(\boldsymbol{\eta}_{1}^{i})\dfrac{\pi(\boldsymbol{\eta}_{1}^{i})}{\pi(\boldsymbol{\eta})}-\alpha_{i}(\boldsymbol{\eta}),\\
		B_{i}(\boldsymbol{\eta)} & =\beta_{i}(\boldsymbol{\eta}_{2}^{i})\dfrac{\pi(\boldsymbol{\eta}_{2}^{i})}{\pi(\boldsymbol{\eta})}-\beta_{i}(\boldsymbol{\eta}),\\
		\Lambda_{i}(\boldsymbol{\eta}) & =\lambda_{i}(\boldsymbol{\eta}_{3}^{i})\dfrac{\pi(\boldsymbol{\eta}_{3}^{i})}{\pi(\boldsymbol{\eta})}-\lambda_{i}(\boldsymbol{\eta}).
	\end{align}
	Because of the periodicity, the master equation $\eqref{mas_eq2}$ is satisfied if one has 
	\begin{equation}\label{local_divergence}
		A_{i}(\boldsymbol{\eta})+B_{i}(\boldsymbol{\eta})+\Lambda_{i}(\boldsymbol{\eta})=\Phi_{i}(\boldsymbol{\eta})-\Phi_{i+1}(\boldsymbol{\eta}).
	\end{equation}
	This condition applies to all configurations of the system, characterized by a set of functions $\Phi_{i}(\boldsymbol{\eta})$ that satisfy the condition $\Phi_{N+1}(\boldsymbol{\eta})=\Phi_{1}(\boldsymbol{\eta})$. The lattice divergence condition \eqref{local_divergence} can be regarded as Noether's theorem in discrete form.
	
	One can equivalently rewrite equation \eqref{local_divergence} using the headway variables introduced in Appendix \ref{hwp}. This allows us to determine the form of the functions $\Phi_i$ and then derive constraints as in equations \eqref{rate_1}--\eqref{rate_3}. For further details of the proof, refer to Appendix \ref{hwp}.
	
	\begin{remark}
		It turns out that our model is indeed a generalization of two models: the RNA Polymerase with minimal interaction model proposed by Belitsky and Schütz \cite{Belitsky2019-1, Belitsky2019-2}, and the cooperation exclusion process considered in \cite{Antal2000}. These processes are discussed in Appendix \ref{consequences}.
	\end{remark}

	\section{Average stationary current and velocity}\label{current_velocity}
	We will utilize the grand canonical ensemble \eqref{grand_can_ensem} for convenience. To compute the average stationary and velocity of particles and buses, we first, following the approach in \cite{Belitsky2019-1,Ngoc2024}, calculate the average densities $\rho^1 := \langle \delta_{s_i,1} \rangle$ and $\rho^2 := \langle \delta_{s_i,2} \rangle$ of particles in states 1 and 2, respectively. Here, $\langle\cdot \rangle$ denotes taking the expectation with respect to the distribution \eqref{grand_can_ensem}. Given that the form of the grand canonical ensemble \eqref{grand_can_ensem} matches that in \cite{Belitsky2019-1}, we can easily obtain the following expressions:
	\begin{equation}\label{state-density}
		\rho^1 = \frac{1}{1+x} \rho;\quad \rho^2 = \frac{x}{1+x} \rho,
	\end{equation}
	where $\rho$ represents the particle density of the two types, i.e., $\rho = \rho^1 + \rho^2$.
	
	Additionally, as mentioned in \cite{Belitsky2019-1}, one can calculate the average excess density of type 1 over type 2. This serves as a straightforward measure for defining particle distribution. The average excess is as follows:
	\begin{equation}
		\sigma := \dfrac{\left< N^1 \right> - \left< N^2 \right>}{L} = \dfrac{1-x}{1+x}\rho.
	\end{equation}
	
	Furthermore, the average dwell time that a particle spends in states 1 and 2, defined as $\tau^\alpha = \rho^\alpha/\rho$ where $\alpha \in \{1,2\}$, can be obtained:
	\begin{equation}\label{dwelltime}
		\tau^1 = \dfrac{1}{1+x}, \ \ \tau^2 = \dfrac{x}{1+x}.
	\end{equation}
	Again, it is worth noting that while the forms of certain expressions in this work resembles those in \cite{Belitsky2019-1,Ngoc2024}, they differ due to variations in parameter values resulting from disparities in model settings.
	\subsection{Average current and velocity of particles in DBRM}
	The average stationary current $j$ of particles, quantified as the average number of particles traversing a site per unit time, reflects the collective motion of these particles. Additionally, the average speed $\nu$ of particles is another characteristic of collective movement, which is linked to $j$ through the equation
	\begin{equation}
		j =\rho \nu.
	\end{equation}

	Instead of using the form of rates \eqref{rate_1}--\eqref{rate_2} to compute $j$ and $\nu$, it is equivalent to use the rates in headway form \eqref{rate1_h}--\eqref{rate2_h} for convenience. The advantage is that it incorporates the headway distribution \eqref{headway_distribution} into the computation. Namely, one has
	\begin{align}
		&j =  \langle \delta_{s_i,2}[\alpha^\star(1-\theta_{i-1}^0) +\alpha^\star(1+\alpha^{1\star}\theta_{i-1}^0)](1-\theta_{i}^0) \rangle \nonumber \\ &+\langle\delta_{s_i,1}[\beta^\star(1-\theta_{i-1}^0)+\beta^\star(1+\beta^{1\star}\theta_{i-1}^0)](1-\theta_{i}^0) \rangle. 
	\end{align}
	Here, again, $\langle\cdot \rangle$ denotes taking the expectation with respect to the distribution \eqref{grand_can_ensem}. By employing the factorization property of the grand canonical ensemble \eqref{grand_can_ensem} and the identities \eqref{state-density}, one can derive the following:
	\begin{align}
		&j = \dfrac{x}{1+x}\rho\big(\alpha^\star(1-\langle\theta_{i-1}^0\rangle)+\alpha^\star(1+\alpha^{1\star}\langle\theta_{i-1}^0\rangle)\big)(1-\langle\theta_{i}^0\rangle)\nonumber\\
		&+\dfrac{1}{1+x}\rho\big(\beta^\star(1-\langle\theta_{i-1}^0\rangle)+\beta^\star(1+\beta^{1\star}\langle\theta_{i-1}^0\rangle)\big)(1-\langle\theta_{i}^0\rangle).
	\end{align}
	
	One can compute $\langle \theta_{i}^0\rangle$ using the headway distribution \eqref{headway_distribution}. Notice that $\langle \theta_{i}^0\rangle$ does not depend on the index $i$ due to translation invariance. Thus, at this stage, we have all the necessary components to plot graphs illustrating the average current (Figs. \ref{fig:j1} and \ref{fig:j2}) and velocity (Figs. \ref{fig:v1} and \ref{fig:v2}) of particles. We examine the impact of neighboring-effect parameters $\alpha^{1\star}$, $\beta^{1\star}$, and the rate $\lambda^\star$ on the average current and velocity of particles. As mentioned previously, we assume $\alpha^\star > \beta^\star$ to model a scenario where a bus, not needing to pick up passengers at a bus stop, can travel faster. Additionally, we set $\alpha^{1\star}$ and $\beta^{1\star}$ to be negative to align with the observation in O'Loan et al.'s BRM. We focus solely on the scenario where $\alpha^{1\star} \leq \beta^{1\star}$, as the inverse case $\alpha^{1\star} \geq \beta^{1\star}$ does not significantly differ in its effect on the quantities.
	
\begin{widetext}

	\begin{figure}[H]
		\centering
		\subfloat[$\beta^\star = 0.1.$	\label{fig:j11}]{\includegraphics[width=5cm]{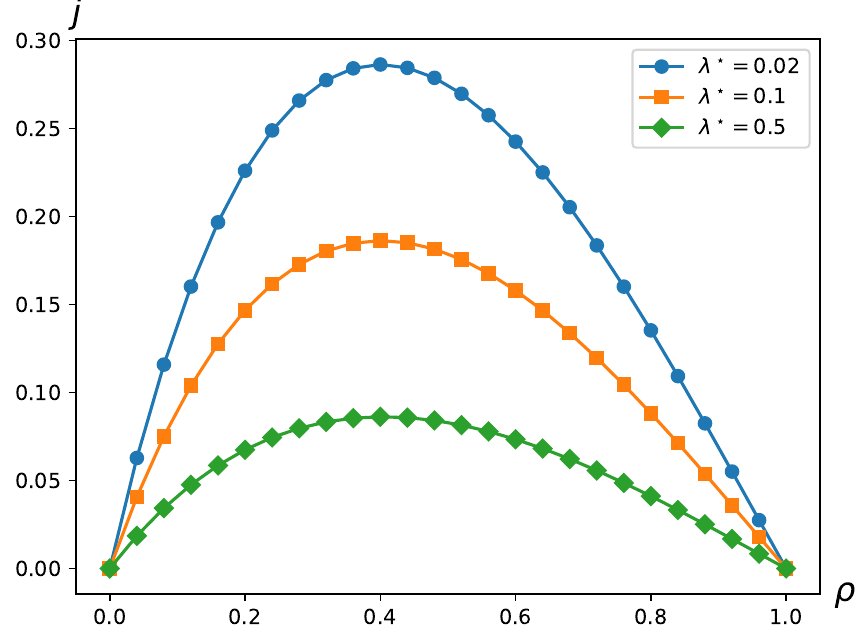}}
		\hspace{1.5cm}
		\subfloat[$\beta^\star = 0.5.$
		\label{fig:j12}]{\includegraphics[width=5cm]{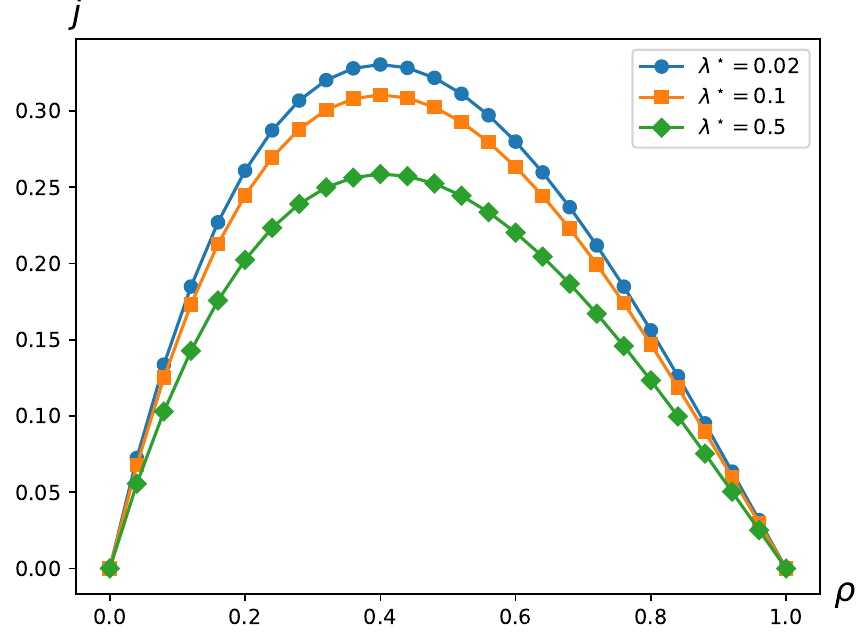}}\\
		\caption{Average current $j$ as a function of particle density $\rho$: $\alpha^\star = 1$, $\alpha^{1\star} = -0.2$, $\beta^{1\star} = -0.1$. The values of $\beta^\star$ are $0.1$ on the left and $0.5$ on the right. Different values of $\lambda^{\star}$ are also included: $0.02$, $0.1$, and $0.5$.}
		\label{fig:j1}
	\end{figure}

	\begin{figure}[H]
		\centering
		\subfloat[$\beta^\star = 0.1.$
		\label{fig:j21}]{\includegraphics[width=5cm]{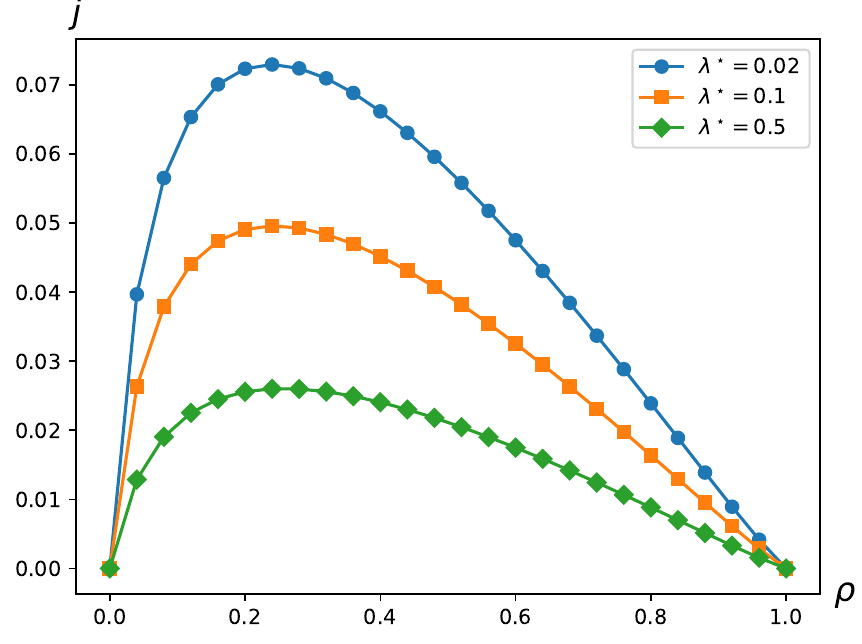}}
		\hspace{1.5cm}
		\subfloat[$\beta^\star = 0.5.$
		\label{fig:j22}]{\includegraphics[width=5cm]{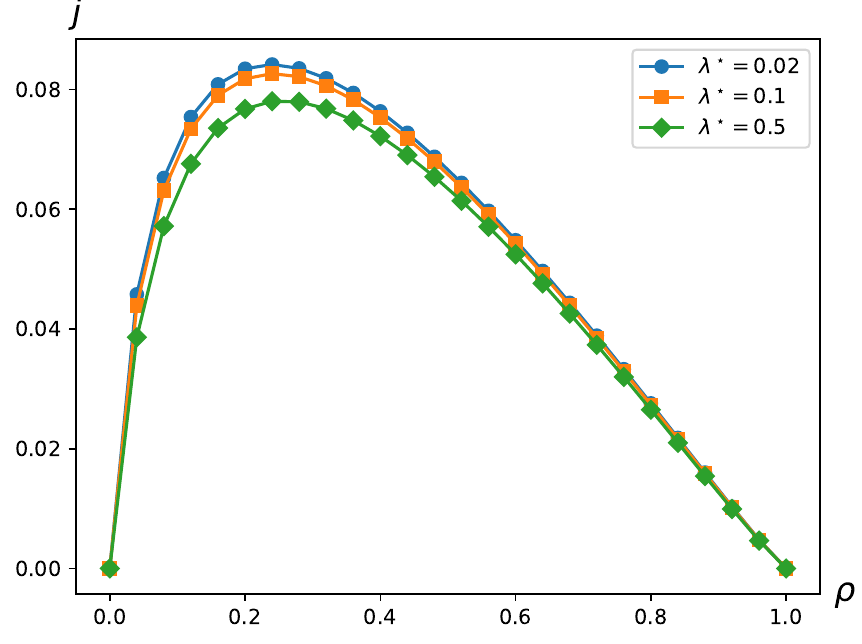}}\\
		\caption{ Average current $j$ as a function of particle density $\rho$: $\alpha^\star = 1$, $\alpha^{1\star} = -0.9$, $\beta^{1\star} = -0.8$. The values of $\beta^\star$ are $0.1$ on the left and $0.5$ on the right. Different values of $\lambda^{\star}$ are also included: $0.02$, $0.1$, and $0.5$.}
		\label{fig:j2}
	\end{figure}

	\begin{figure}[h]
		\centering
		\subfloat[$\beta^\star = 0.1.$
		\label{fig:v11}]{\includegraphics[width=5cm]{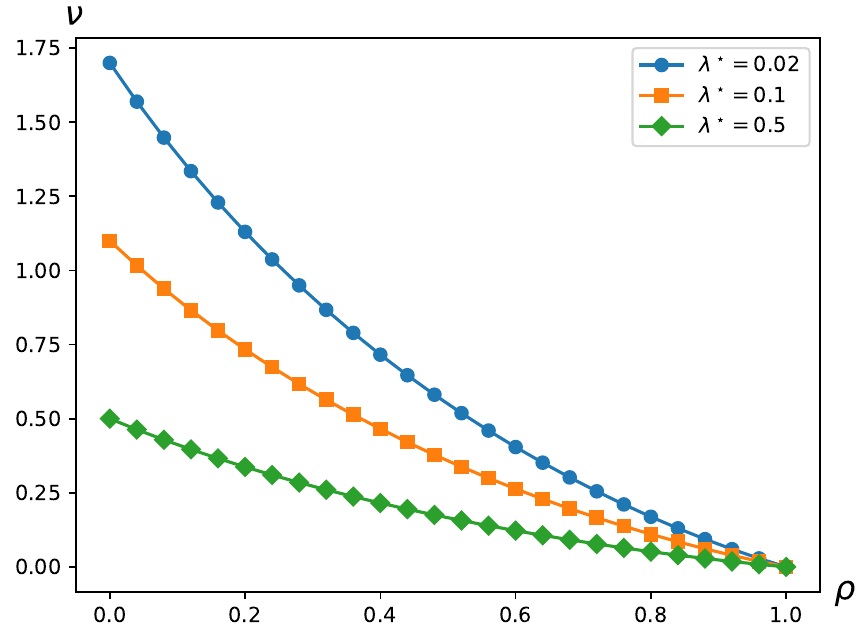}}
		\hspace{1.5cm}
		\subfloat[$\beta^\star = 0.5.$
		\label{fig:v12}]{\includegraphics[width=5cm]{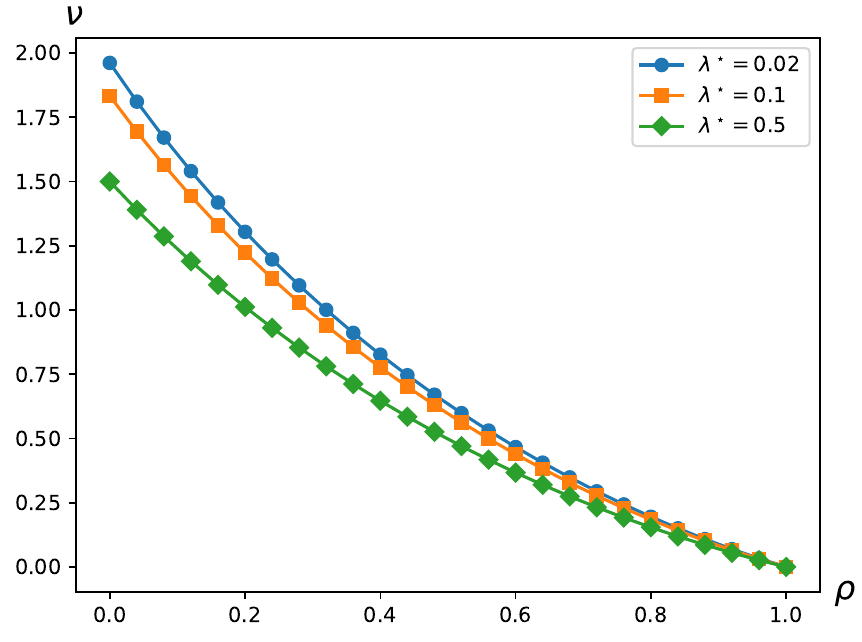}}\\
		\caption{Average velocity $\nu$ as a function of particle density $\rho$: $\alpha^\star = 1$, $\alpha^{1\star} = -0.2$, $\beta^{1\star} = -0.1$. The values of $\beta^\star$ are $0.1$ on the left and $0.5$ on the right. Different values of $\lambda^{\star}$ are also included: $0.02$, $0.1$, and $0.5$.}
		\label{fig:v1}
	\end{figure}

	\begin{figure}[h]
		\centering
		\subfloat[$\beta^\star = 0.1.$
		\label{fig:v21}]{\includegraphics[width=5cm]{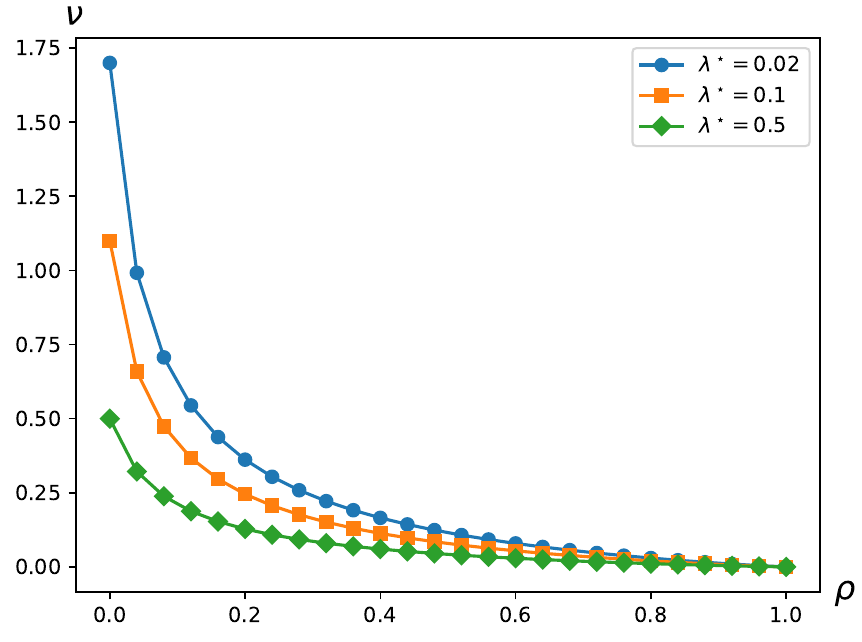}}
		\hspace{1.5cm}
		\subfloat[$\beta^\star = 0.5.$
		\label{fig:v22}]{\includegraphics[width=5cm]{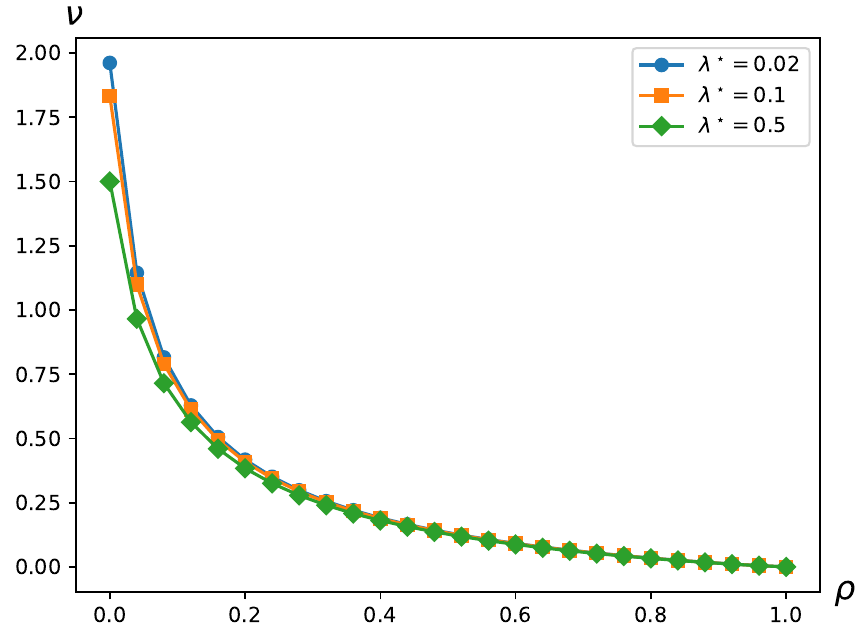}}\\
		\caption{Average velocity $\nu$ as a function of particle density $\rho$: $\alpha^\star = 1$, $\alpha^{1\star} = -0.9$, $\beta^{1\star} = -0.8$. The values of $\beta^\star$ are $0.1$ on the left and $0.5$ on the right. Different values of $\lambda^{\star}$ are also included: $0.02$, $0.1$, and $0.5$.}
		\label{fig:v2}
	\end{figure}

\end{widetext}

	\subsection{Average current and velocity of buses in BRM}
	Since the average stationary current $j$ and velocity $\nu$ of the particles obtained ,  thus  the average current $j_b$ and velocity $\nu_b$ of buses with density $\rho_b = 1 - \rho$ can be computed as follows
	\begin{equation}
		\begin{cases}
			j_b(\rho_b) = j(1-\rho_b),\\
			\nu_b(\rho_b) = j_b(\rho_b)/\rho_b.
		\end{cases}
	\end{equation}
	Thus, one can plot the average current and velocity of the buses, see Figs. \ref{fig:jb1} and \ref{fig:jb2} for the current and Figs. \ref{fig:vb1} and \ref{fig:vb2} for the velocity.

\begin{widetext}

	\begin{figure}[H]
		\centering
		\subfloat[$\beta^\star = 0.1.$
		\label{fig:jb11}]{\includegraphics[width=5cm]{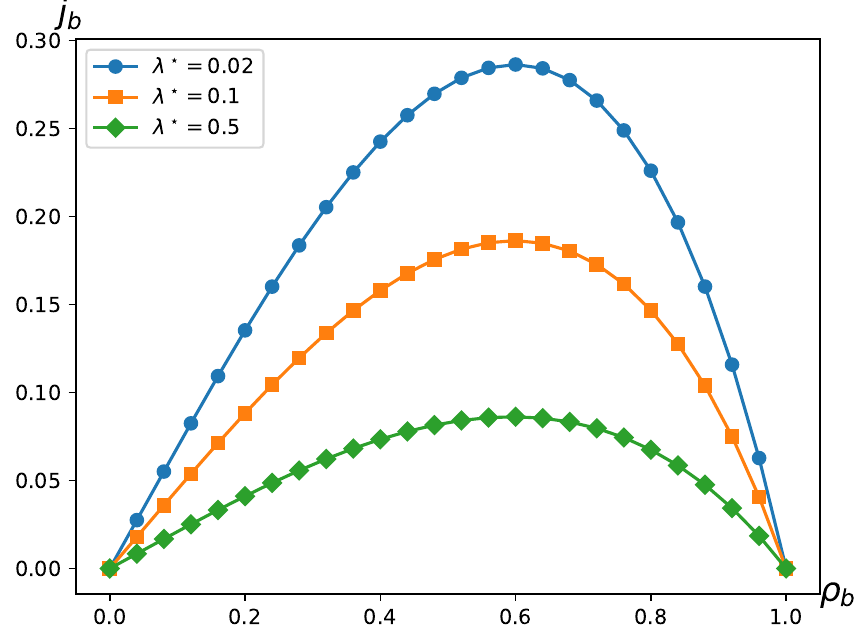}}
		\hspace{1.5cm}
		\subfloat[$\beta^\star = 0.5.$
		\label{fig:jb12}]{\includegraphics[width=5cm]{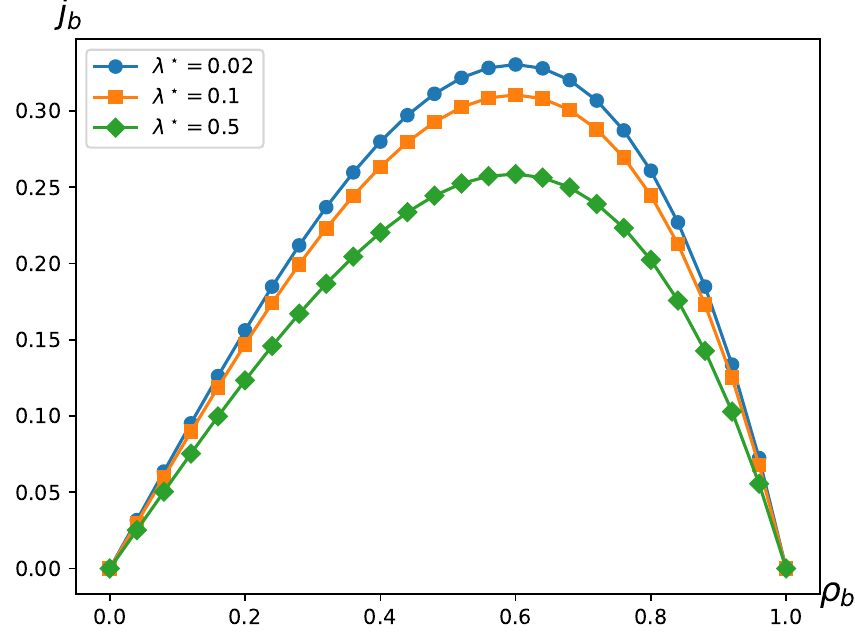}}\\
		\caption{Average current $j_b$ as a function of bus density $\rho_b$: $\alpha^\star = 1$, $\alpha^{1\star} = -0.2$, $\beta^{1\star} = -0.1$. The values of $\beta^\star$ are $0.1$ on the left and $0.5$ on the right. Different values of $\lambda^{\star}$ are also included: $0.02$, $0.1$, and $0.5$.}
		\label{fig:jb1}
	\end{figure}

	\begin{figure}[H]
		\centering
		\subfloat[$\beta^\star = 0.1.$
		\label{fig:jb21}]{\includegraphics[width=5cm]{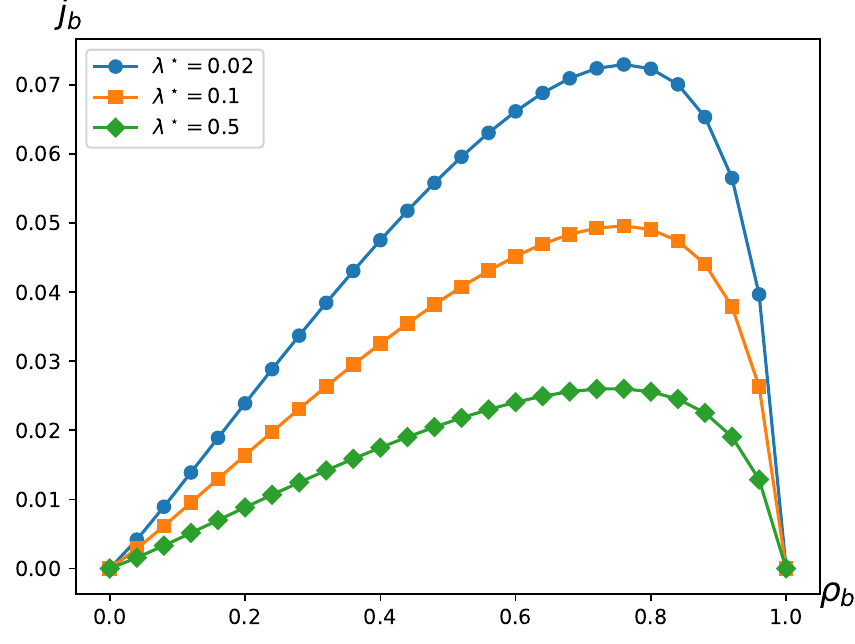}}
		\hspace{1.5cm}
		\subfloat[$\beta^\star = 0.5.$
		\label{fig:jb22}]{\includegraphics[width=5cm]{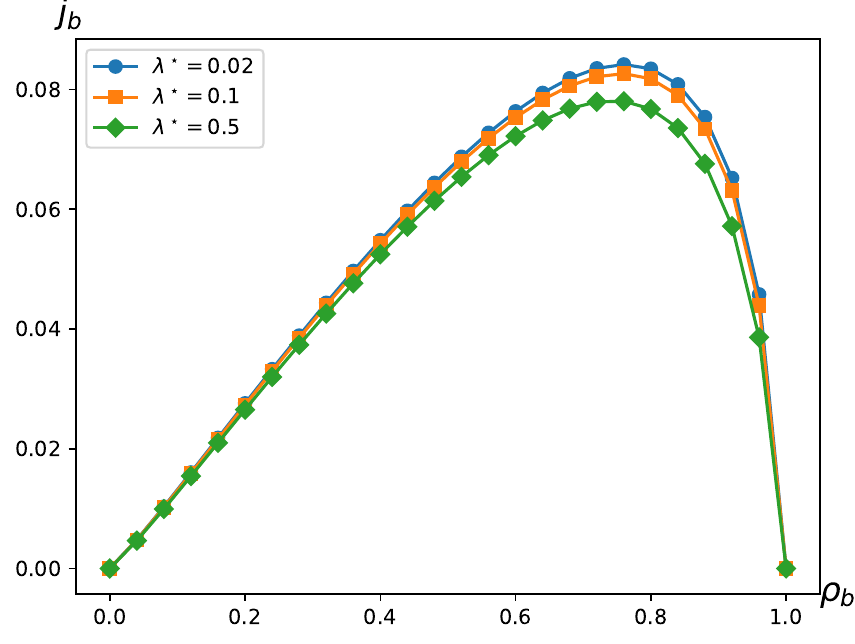}}\\
		\caption{Average current $j_b$ as a function of bus density $\rho_b$: $\alpha^\star = 1$, $\alpha^{1\star} = -0.9$, $\beta^{1\star} = -0.8$. The values of $\beta^\star$ are $0.1$ on the left and $0.5$ on the right. Different values of $\lambda^{\star}$ are also included: $0.02$, $0.1$, and $0.5$.}
		\label{fig:jb2}
	\end{figure}

	\begin{figure}[H]
		\centering
		\subfloat[$\beta^\star = 0.1.$
		\label{fig:vb11}]{\includegraphics[width=5cm]{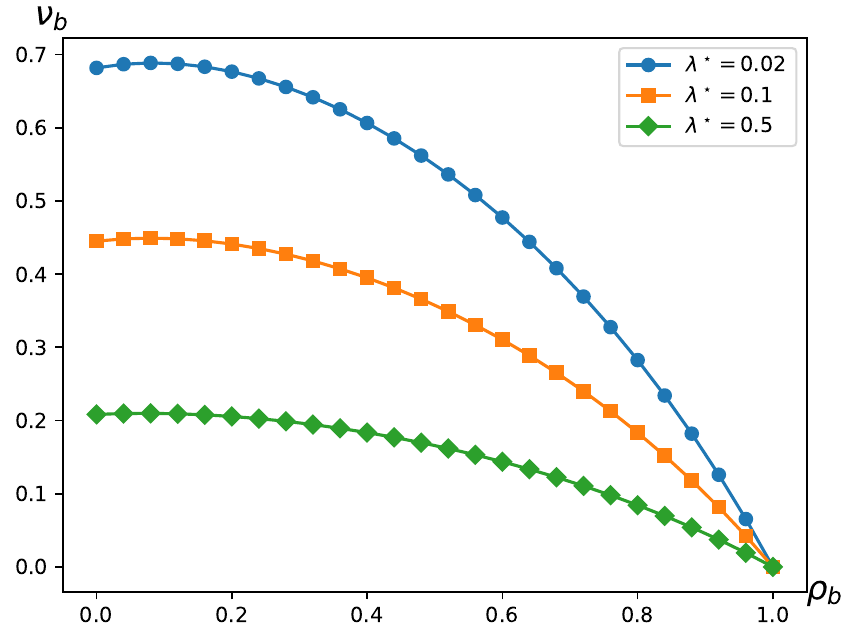}}
		\hspace{1.5cm}
		\subfloat[$\beta^\star = 0.5.$
		\label{fig:vb12}]{\includegraphics[width=5cm]{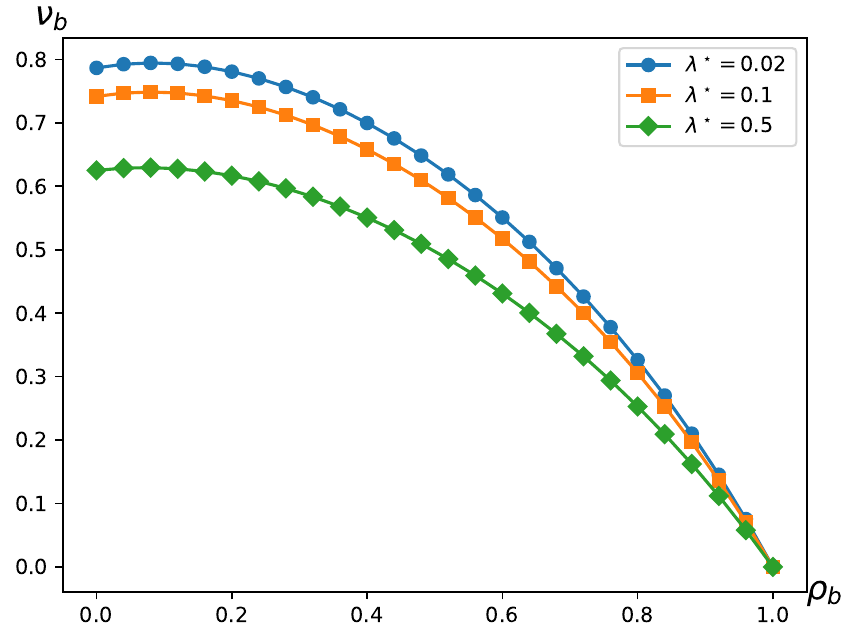}}\\
		\caption{Average velocity $\nu_b$ as a function of bus density $\rho_b$: $\alpha^\star = 1$, $\alpha^{1\star} = -0.2$, $\beta^{1\star} = -0.1$. The values of $\beta^\star$ are $0.1$ on the left and $0.5$ on the right. Different values of $\lambda^{\star}$ are also included: $0.02$, $0.1$, and $0.5$.}
		\label{fig:vb1}
	\end{figure}

	\begin{figure}[H]
		\centering
		\subfloat[$\beta^\star = 0.1.$
		\label{fig:vb21}]{\includegraphics[width=5cm]{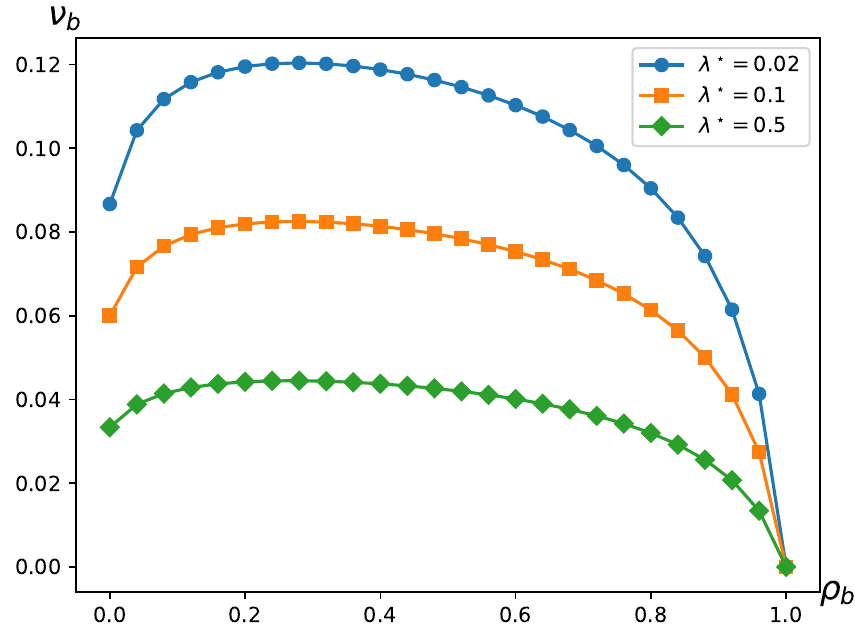}}
		\hspace{1.5cm}
		\subfloat[$\beta^\star = 0.5.$
		\label{fig:vb22}]{\includegraphics[width=5cm]{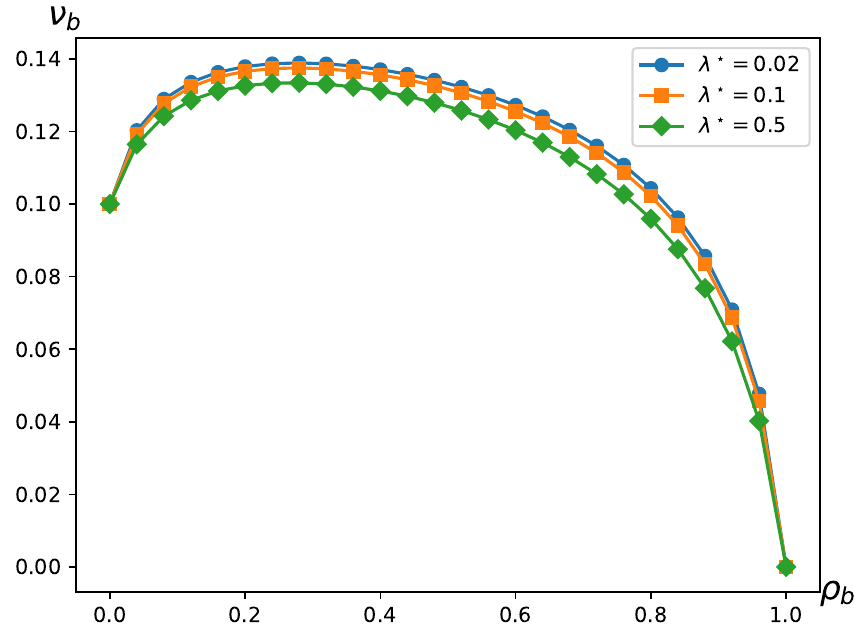}}\\
		\caption{Average velocity $\nu_b$ as a function of bus density $\rho_b$: $\alpha^\star = 1$, $\alpha^{1\star} = -0.9$, $\beta^{1\star} = -0.8$. The values of $\beta^\star$ are $0.1$ on the left and $0.5$ on the right. Different values of $\lambda^{\star}$ are also included: $0.02$, $0.1$, and $0.5$.}
		\label{fig:vb2}
	\end{figure}

\end{widetext}		
	
	\section{Discussion}\label{discussion}
	It is evident that when $\beta^\star$ is large, the current and velocity of particles, as well as of buses, are quite similar, as shown in the left panels of Figs. \ref{fig:v1}, \ref{fig:v2}, \ref{fig:vb1}, and \ref{fig:vb2}. This suggests that when buses pick up passengers quickly, the velocity and current are not significantly affected. Moreover, when comparing the left and right panels of each figure, one can observe that the current and velocity with a higher value of $\beta^\star$ are greater than those with a smaller value. Additionally, it is evident that as the value of $\lambda^\star$ decreases, both the current and velocity increase. This can be explained easily, as a smaller $\lambda^\star$ implies fewer passengers arriving at bus stops, resulting in less time spent by buses picking them up.

	Let us now consider the impact of parameters $\alpha^{1\star}$ and $\beta^{1\star}$ on the average stationary current and velocity of particles, which we shall refer to as the neighboring effect. We classify this influence as either strong or weak, depending on whether $\alpha^{1\star}$ and $\beta^{1\star}$ approximate -1 or 0, respectively. When the effect is not strong, indicating that $\alpha^{1\star}$ and $\beta^{1\star}$ are close to 0, our model behaves similarly to the one described in \cite{OLoan1998-1, OLoan1998-2}. Specifically, let us examine the current and velocity shown in Figs. \ref{fig:j1} and \ref{fig:v1} for DBRM, and in Figs. \ref{fig:jb1} and \ref{fig:vb1} for BRM. The average velocity of buses decreases to 0 as the density of buses increases to 1, as depicted in Fig. \ref{fig:vb1}. It is worth noting that Fig. \ref{fig:vb12} bears resemblance to Fig. 4 in \cite{OLoan1998-1}. Additionally, regarding the average current of particles in DBRM, Fig. \ref{fig:j1} shares similarities with Fig. 18 in \cite{OLoan1998-1}, indicating the similarity of our model to the original.
	
	When the neighboring effect is strong, meaning $\alpha^{1\star}$ and $\beta^{1\star}$ are close to -1, the behavior of average current and velocity of particles, and hence buses, becomes particularly intriguing. This phenomenon is depicted in Figs. \ref{fig:j2} and \ref{fig:v2} for particles in DBRM, and Figs. \ref{fig:jb2} and \ref{fig:vb2} for buses in BRM. Notably, the current (Fig. \ref{fig:j2}) of particles in DBRM rapidly reaches a maximum at very low densities, then decreases to zero as the density goes to 1. Similarly, the average velocity of particles in DBRM drops rapidly at low density and then decreases gradually to 0 in areas with higher density, as shown in Fig. \ref{fig:v2}. In terms of the current and velocity of buses in BRM under strong neighboring-effect parameters, the current reaches a maximum at very high density and after that drops quickly to zero, as depicted in Fig. \ref{fig:jb2}. Interestingly, in the velocity case (Fig. \ref{fig:vb2}), the velocity increases even at very low density under strong neighboring-effect parameters. This behavior contrasts with weaker neighboring-effect scenarios, as observed in the original BRM. After reaching a maximum, it decreases slowly to 0.

	\section{Conclusion}\label{Summary}
	
	We have presented an exactly solvable DBRM, which allows us to investigate the BRM proposed by O'Loan et al. Our model differs from the original BRM by introducing neighboring-effect parameters that describe the observation in the original model: a bus moves slower when it is farther from the leading bus.
	
	While our findings closely mirror those of O'Loan et al. when the neighboring effect is weak, there is a notable departure when the effect is strong. In this case, the behavior of the average velocity of buses becomes intriguing, presenting a phenomenon not observed in the original BRM. Specifically, even at low bus density, the average velocity of buses can increase as the density rises.
	
	\section*{Appendices}
	\appendix
	\renewcommand{\thesection}{\Alph{section}}
	\section{Mapping to headway process }\label{hwp}
	In addition to defining a system's configuration with its position and state vectors $\textbf{x} = (x_1,...,x_N)$ and $\textbf{s} = (s_1,...,s_N)$ respectively, we can also specify it using the state vector alongside the headway vector $\textbf{m} = (m_1,...,m_N)$. Here, $m_i$ denotes the count of empty sites between the $i$-th and $(i+1)$-th particles, where $m_i = x_{i+1} - (x_i + 1) \mod L$.
	
	To represent a headway of length $r$, rather than using $\delta_{m_i,r}$, we introduce the more compact notation $\theta_{i}^r$. Specifically, we define
	\begin{equation}\label{newvar}
		\theta_{i}^r := \delta_{m_i, r} = \delta_{x_{i+1}, x_i + 1 +r},\ \ \text{for } r \in \{0,1,2,...\}.
	\end{equation}
	Regarding the headway variable, the index $i$ is considered modulo $N$, thus implying $\theta_0^r \equiv \theta_N^r$.
	
	Due to steric hard-core repulsion, a forward translocation from $x_{i}$ to $x_{i}+1$ of the $i$-th particle  corresponds to the transition $(m_{i-1},m_{i})\to(m_{i-1}+1,m_{i}-1)$, provided that $m_{i}>0$. When expressed in the new stochastic variables $\boldsymbol{\zeta}=(\textbf{m},\textbf{s})$, where $\textbf{m}$ denotes the distance vector and $\textbf{s}$ represents the state vector, the invariant distribution \eqref{inv_meas} can be reformulated as
	\begin{equation}\label{inv_meas_2}
		\tilde{\pi}(\boldsymbol{\zeta}) = \dfrac{1}{\tilde{Z}} \prod_{i=1}^{N}x^{3/2-s_i}y^{-\theta_{i}^0},
	\end{equation} 
	where $\tilde{Z}$ denotes the partition function. Moreover, the transition rates \eqref{rate_1}--\eqref{rate_3} can be reformulated as
	\begin{align}
		&	\tilde{\alpha}_{i}(\boldsymbol{\zeta})  =\alpha^{\star}\delta_{s_{i},2}(1+\alpha^{1\star}\theta_{i-1}^{0})(1-\theta_{i}^{0}),\label{rate1_h}\\
		&	\tilde{\beta}_{i}(\boldsymbol{\zeta})  =\beta^{\star}\delta_{s_{i},1}(1+\beta^{1\star}\theta_{i-1}^{0})(1-\theta_{i}^{0}),\label{rate2_h}\\
		&	\tilde{\lambda}_{i}(\boldsymbol{\zeta})  =\lambda^{\star}\delta_{s_{i},2}(1+\lambda^{1\star}\theta_{i-1}^{0}+\lambda^{\star1}\theta_{i}^{0}+\lambda^{1\star1}\theta_{i-1}^0\theta_{i}^0).\label{rate3_h}
	\end{align}

	Given a configuration $\boldsymbol{\zeta}$, let $\boldsymbol{\zeta}^{i,1}$ and $\boldsymbol{\zeta}^{i,2}$ represent the configurations resulting from the forward translocation when the state of particle $i$ is 1 and 2, respectively. Additionally, let $\boldsymbol{\zeta}^{i,3}$ denote the configuration leading to $\boldsymbol{\zeta}$ before a passenger arrives at site $x_i$. Therefore, the configurations $\boldsymbol{\zeta}^{i,1}$, $\boldsymbol{\zeta}^{i,2}$, and $\boldsymbol{\zeta}^{i,3}$ are defined by the distance and state vectors as follows:
	\begin{align}
		&m^{i,1}_j := m_j + \delta_{j,i-1} - \delta_{j,i};\ \ s^{i,1}_j := s_j + (3-2s_j) \delta_{j,i},\\
		&m^{i,2}_j := m_j + \delta_{j,i-1} - \delta_{j,i};\ \  s^{i,2}_j := s_j,\\
		&m_j^{i,3} := m_j;\ \ s_j^{i,3}  := s_j + (3-2s_j)\delta_{j,i}.
	\end{align}
	This yields the master equation for the headway process
	\begin{equation}\label{mas_eq3}
		\dfrac{d\mathbb{P}(\boldsymbol{\zeta},t)}{dt} = \sum_{i=1}^{N}H_i(\boldsymbol{\zeta},t)
	\end{equation}
	with
	\begin{align}
		\begin{split}
		H_i(\boldsymbol{\zeta},t) &=\ \tilde{\alpha}_i(\boldsymbol{\zeta}^{i,1})\mathbb{P}(\boldsymbol{\zeta}^{i,1},t) - \tilde{\alpha}_i(\boldsymbol{\zeta})\mathbb{P}(\boldsymbol{\zeta},t)\\ 
			&+ \tilde{\beta}_i(\boldsymbol{\zeta}^{i,2})\mathbb{P}(\boldsymbol{\zeta}^{i,2},t) - \tilde{\beta}_i(\boldsymbol{\zeta})\mathbb{P}(\boldsymbol{\zeta},t) 
			 \\&+ \tilde{\lambda}_i(\boldsymbol{\zeta}^{i,3})\mathbb{P}(\boldsymbol{\zeta}^{i,3},t) - \tilde{\lambda}_i(\boldsymbol{\zeta})\mathbb{P}(\boldsymbol{\zeta},t) 
		\end{split}
	\end{align}
	where
	\begin{align}
		\tilde{\alpha}_{i}(\boldsymbol{\zeta}^{i,1}) & =\alpha^{\star}\delta_{s_{i},2}(1+\alpha^{\star1}\theta_{i-1}^{1})(1-\theta_{i-1}^{0}),\\
		\tilde{\beta}_{i}(\boldsymbol{\zeta}^{i,2}) & =\beta^{\star}\delta_{s_{i},2}(1+\beta^{\star1}\theta_{i-1}^{1})(1-\theta_{i-1}^{0}),\\
		\tilde{\lambda}_{i}(\boldsymbol{\zeta}^{i,3}) & =\lambda^{\star}\delta_{s_{i},1}(1+\lambda^{1\star}\theta_{i-1}^{0}+\lambda^{\star1}\theta_{i}^{0}+\lambda^{1\star1}\theta_{i-1}^0\theta_{i}^0).
	\end{align}
	Thus, at equilibrium, the master equation \eqref{mas_eq3} yields
	\begin{align}
		\sum_{i=1}^{N}\bigg(&\tilde{\alpha}_i(\boldsymbol{\zeta}^{i,1})\tilde{\pi}(\boldsymbol{\zeta}^{i,1}) 
		+ \tilde{\beta}_i(\boldsymbol{\zeta}^{i,2})\tilde{\pi}(\boldsymbol{\zeta}^{i,2}) + \tilde{\lambda}_i(\boldsymbol{\zeta}^{i,3})\tilde{\pi}(\boldsymbol{\zeta}^{i,3}) \nonumber \\
		&- \big(\tilde{\alpha}_i(\boldsymbol{\zeta}) + \tilde{\beta}_i(\boldsymbol{\zeta})  + \tilde{\lambda}_i(\boldsymbol{\zeta})\big)\tilde{\pi}(\boldsymbol{\zeta})\bigg) = 0.\label{mas_eq4}
	\end{align}
	
	Before writing the local divergence condition for the headway process, which is equivalent to \eqref{local_divergence}, we establish the notation
	\begin{align}
		\tilde{A}_i(\boldsymbol{\zeta}) &= \tilde{\alpha}_i(\boldsymbol{\zeta}^{i,1})\dfrac{\tilde{\pi}(\boldsymbol{\zeta}^{i,1})}{\tilde{\pi}(\boldsymbol{\zeta})} -\tilde{\alpha}_i(\boldsymbol{\zeta}),\\
		\tilde{B}_i(\boldsymbol{\zeta}) &= \tilde{\beta}_i(\boldsymbol{\zeta}^{i,2})\dfrac{\tilde{\pi}(\boldsymbol{\zeta}^{i,2})}{\tilde{\pi}(\boldsymbol{\zeta})} -\tilde{\beta}_i(\boldsymbol{\zeta}),\\
		\tilde{\Lambda}_i(\boldsymbol{\zeta}) &= \tilde{\lambda}_i(\boldsymbol{\zeta}^{i,3})\dfrac{\tilde{\pi}(\boldsymbol{\zeta}^{i,3})}{\tilde{\pi}(\boldsymbol{\zeta})} -\tilde{\lambda}_i(\boldsymbol{\zeta}).
	\end{align}
	One has
		\begin{align}
			&\theta_j^p(\boldsymbol{\zeta}^{i,1}) = \delta_{m_j+\delta_{j,i-1}-\delta_{j,i},p} = \theta_j^{p-\delta_{j,i-1}+\delta_{j,i}}(\boldsymbol{\zeta}),\\
			& \delta_{s_i^{i,1},\alpha} =\delta_{s_i,3-\alpha};\\
			&\theta_j^p(\boldsymbol{\zeta}^{i,2}) = \delta_{m_j+\delta_{j,i-1}-\delta_{j,i},p} = \theta_j^{p-\delta_{j,i-1}+\delta_{j,i}}(\boldsymbol{\zeta}),\\
			& \delta_{s_i^{i,2},\alpha} =\delta_{s_i,\alpha};\\
			&\theta_j^p(\boldsymbol{\zeta}^{i,3}) = \theta_j^{p}(\boldsymbol{\zeta}) \text{ and } \delta_{s_i^{i,3},\alpha} =\delta_{s_i,3-\alpha}.
		\end{align}
	Thus, one gets
	\begin{align}
		\dfrac{\tilde{\pi}(\boldsymbol{\zeta}^{i,1})}{\tilde{\pi}(\boldsymbol{\zeta})} & =x^{3-2s_{i}}y^{\theta_{i-1}^{0}-\theta_{i-1}^{1}+\theta_{i}^{0}},\\
		\dfrac{\tilde{\pi}(\boldsymbol{\zeta}^{i,2})}{\tilde{\pi}(\boldsymbol{\zeta})} & =y^{\theta_{i-1}^{0}-\theta_{i-1}^{1}+\theta_{i}^{0}},\\
		\dfrac{\tilde{\pi}(\boldsymbol{\zeta}^{i,3})}{\tilde{\pi}(\boldsymbol{\zeta})} & =x^{3-2s_{i}}.
	\end{align}
	
	Similarly to \eqref{local_divergence}, the master equation $\eqref{mas_eq4}$ is satisfied if one requires
	\begin{equation}\label{Noether}
		\tilde{A}_i + \tilde{B}_i + \tilde{\Lambda}_i  = \tilde{\Phi}_{i-1} - \tilde{\Phi}_{i},
	\end{equation}
	where $\tilde{\Phi}_i$ is of the form $\tilde{\Phi}_i = (a+ b\theta_{i}^0 + c\theta_{i}^1)(\delta_{s_i,1}+\delta_{s_i,2}) = a+ b\theta_{i}^0 + c\theta_{i}^1$. Note that $\tilde{\Phi}_i$ must conform to this form, as $\tilde{A}_i$, $\tilde{B}_i$, and $\tilde{\Lambda}_i$ are contingent upon the state of particle $i$ and variables $\theta_{i-1}^0$, $\theta_{i-1}^1$, $\theta_{i}^0$, $\theta_{i}^1$, all of which belong to the set $\{0,1\}$. By examining all possible cases of \eqref{Noether}, one can initially obtain
	
	\begin{align}
		b & = \dfrac{-\alpha^{\star}(1+\lambda^{1\star})-x\beta^{\star}(1+\beta^{1\star})}{1+x},\label{value_b}\\
		c & = 0,\label{value_c}
	\end{align}
	and then one gets the stationary conditions \eqref{cond1}--\eqref{cond2}. 

	\section{Partition function and headway distribution}
	
	Just like in previous studies \cite{Belitsky2019-1,Belitsky2019-2,Ngoc2024_2}, instead using measure \eqref{inv_meas_2}, we adopt the grand-canonical ensemble for ease of analysis, as defined by
	\begin{equation}\label{grand_can_ensem}
		\hat{\pi}(\boldsymbol{\zeta})=\dfrac{1}{Z_{gc}}\prod_{i=1}^{N}x^{-3/2+s_{i}}y^{-\theta_{i}^{0}}z^{m_{i}}
	\end{equation}
	where $Z_{gc}=(Z_{1}Z_{2})^{N}$ with
	\begin{equation}
		Z_{1}=\dfrac{1+(y-1)z}{1-z},\ \ Z_{2}=x^{1/2}+x^{-1/2}.
	\end{equation}
	Here, the fugacity $z$ serves as a fugacity through which the particle density can be determined.
	
	As evident, the partition function $Z_{gc}$ maintains a fixed structure. Therefore, it becomes imperative to establish the well-defined nature of the measure \eqref{grand_can_ensem}. This necessitates demonstrating the equivalence between the two measures \eqref{inv_meas_2} and \eqref{grand_can_ensem}, where a specific value of $z$ corresponding to a given particle density $\rho$ can always be identified, ensuring identical probabilities of a configuration under both measures. It is noteworthy that this undertaking has previously been addressed in \cite{Belitsky2019-1, Belitsky2019-2}, given the resemblance between the grand-canonical ensemble form \eqref{grand_can_ensem} and that presented in the cited papers. As outlined in \cite{Belitsky2019-1,Belitsky2019-2}, the value of $z$ is determined as follows:
	\begin{equation}
		z(\rho,y)=1-\dfrac{1-\sqrt{1-4\rho(1-\rho)(1-y^{-1})}}{2(1-\rho)(1-y^{-1})}\label{value_z}.
	\end{equation}
	
	When dealing with the grand-canonical ensemble \eqref{grand_can_ensem} in the context of the headway process, obtaining the probability distribution for the gap $m_i$ between two particles is straightforward, defined as
	\begin{equation}
		P_h(r) =\hat{\pi}_{gc}(m_i = r),\ \ \text{where } r\in\{0,1,2...\}.
	\end{equation}
	Alternatively, we can represent $P_h(r)$ as $\langle \theta_i^r\rangle$, where the random variables $\theta_i^r$ are defined in equation \eqref{newvar}. Hence, the headway distribution can be formulated as
	\begin{equation}\label{headway_distribution}P_h(r) = 
		\begin{cases}
			\dfrac{1-z}{1+(y-1)z} \ \ \text{ for } r = 0, \\
			yP_h(0)z^r \ \ \text{ for } r \geq 1.
		\end{cases}
	\end{equation}
	
	\section{Limiting cases}\label{consequences}
	\subsection{RNA Polymerase with minimal interaction model}
	If one sets $\alpha_{i}(\boldsymbol{\eta}) = 0$ by letting $\alpha^\star = 0$, the dynamics of our dual bus route model, given by equations \eqref{rate_1}-\eqref{rate_3}, now are valid for only two rates: $\beta_{i}(\boldsymbol{\eta})$ and $\lambda_{i}(\boldsymbol{\eta})$. Thus, our model mathematically reduces to the RNA Polymerase (RNAP) interaction model with minimal interaction, as considered in \cite{Belitsky2019-1, Belitsky2019-2}, which was proposed to mathematically predict the conditions under which collective pushing among RNAPs arises. 
	
	In that model, an RNAP appears in only two distinct polymerization states, namely with or without pyrophosphate (PP$i$, one of the products of hydrolysis) bound to it, corresponding to states 1 and 2. Suppose that $\boldsymbol{\eta}$ is a configuration of RNAP with its corresponding vector of positions $\textbf{x} = (x_1,...,x_N)$ and states $\textbf{s} = (s_1,...,s_N)$ of RNAPs. Thus, the configuration-dependent translocation rate for RNAP $i$ from position $x_i$ to position $x_{i+1}$, $\omega_i(\boldsymbol{\eta})$ (equation (2.8) in \cite{Belitsky2019-1}), and the rate for PP$_i$ release $\kappa_i(\boldsymbol{\eta})$ (equation (2.9) in \cite{Belitsky2019-1}), are as follows:
	\begin{align}
		&\omega_{i}(\boldsymbol{\eta})  =\delta_{s_{i},1}\omega^{\star}(1+\delta_{x_{i-1},x_{i}-1}\omega^{1\star})(1-\delta_{x_{i}+1,x_{i+1}})\label{rate_2'}\\
		\begin{split}
			&\kappa_{i}(\boldsymbol{\eta})  =\delta_{s_{i},2}f^{\star}(1+\delta_{x_{i-1},x_{i}-1}f^{1\star}+\delta_{x_{i}+1,x_{i+1}}f^{\star1}\\
			& + f^{1\star1}\delta_{x_{i-1},x_{i}-1}\delta_{x_{i}+1,x_{i+1}}).\label{rate_3'}
		\end{split}
	\end{align}
	These transition rates share the same forms as $\beta_{i}(\boldsymbol{\eta})$ and $\lambda_{i}(\boldsymbol{\eta})$ in our model, respectively. Of course, the range of parameters appearing in the rates $\omega_i(\boldsymbol{\eta})$ and $\kappa_i(\boldsymbol{\eta})$ must be chosen to adapt to the specific usage in the RNAP model.

	Thus, the conditions \eqref{cond1}--\eqref{cond2} for the corresponding parameters appearing in the rate $\kappa_i(\boldsymbol{\eta})$ become the following:
	\begin{align}
		f^{1\star} &= \dfrac{x}{1+x}(1+\omega^{1\star})-1\label{cond1'}\\
		f^{\star1} &= \dfrac{1}{1+x}(1+\omega^{1\star})-1\\
		f^{1\star1} & = -\omega^{1\star}\label{cond2'}
	\end{align}
	Then, in equilibrium, if the parameters appearing in the rates \eqref{rate_2'}--\eqref{rate_3'} satisfy the constraints \eqref{cond1'}--\eqref{cond2'}, according to \eqref{main_theorem}, on the lattice with periodic conditions, the invariant distribution of the RNAP processes is given by:
	
	\begin{equation}\label{main_theorem1}
		\pi(\boldsymbol{\eta}) = \dfrac{1}{Z}\left(\dfrac{\omega^{\star}}{f^{\star}}\right)^{\sum_{i=1}^{N}-3/2+ s_i}\left( 1+\omega^{1\star} \right)^{-\sum_{i=1}^{N}\delta_{x_{i+1}, x_i+1}}
	\end{equation}
	where $Z$ is the normalized constant.\\

\subsection{Cooperation exclusion process}
	By setting all particles to be distinguishable, i.e., without any degree of freedom, all particles are in the same state, specifically state 2. Thus, in this case, the rate  $\alpha_{i}(\boldsymbol{\eta})$ in \eqref{rate_1} must be equal to 0. Additionally, if one sets the transition rate $\lambda_{i}(\boldsymbol{\eta})$ in \eqref{rate_3} to be 0 by letting $\lambda^\star = 0$, the dynamics of the model \eqref{rate_1}-\eqref{rate_3} for particle $i$ moving from position $x_i$ to $x_{i}+1$ reduce to the following:
	\begin{equation}
		\beta_{i}(\boldsymbol{\eta}) = 	\beta^{\star}(1+\delta_{x_{i-1},x_{i}-1}\beta^{1\star})(1-\delta_{x_{i}+1,x_{i+1}}).
	\end{equation}
	Here, $x_i$ is the position of $i$-th particle in the lattice. 
	
	It turns out that the model with the above dynamics is actually the process called the cooperative exclusion process \cite{Gabel2011}. The dynamics expand upon the TASEP by incorporating interactions between leftmost nearest neighbors. In this particular model, particle motion is governed by the following rules: If the adjacent leftmost nearest neighbor site is vacant, the particle moves to the right with a speed of $\beta^\star$. However, if the site is occupied, the particle moves to the right with a speed of $\beta^\star(1+\beta^{1\star})$. These dynamics can be represented as:
	\begin{equation}\label{model2}
		\begin{aligned}
			& 010 \to 001 \text{ with rate } \beta^\star,\\
			& 110 \to 101 \text{ with rate } \beta^\star(1+\beta^{1\star}).
		\end{aligned}
	\end{equation}
	According to \eqref{main_theorem}, on the ring with periodic boundary conditions, the invariant distribution of the above process is given by:
	\begin{equation}\label{Antal}
		\pi(\boldsymbol{\eta}) = \dfrac{1}{Z}\left( 1+\beta^{1\star}\right)^{\sum_{i=1}^{L}\eta_i\eta_{i+1} }.
	\end{equation}
	Here $Z$ is the partition function. Notice that the above measure \eqref{Antal} is an Ising-like measure \cite{Gabel2011,Katz,Ngoc2024_1,Schutz2017} for exclusion process similar to Ising measure for Ising model \cite{wiki}.


\end{document}